\documentstyle[amssymb]{article}
\newtheorem{th}{Theorem}[section]

\newtheorem{cor}[th]{Corollary}

\begin{document}

\title{Ricci curvature of submanifolds in locally conformal almost cosymplectic
manifolds}
\author{Mukut Mani Tripathi\thanks{%
Partially supported by Brain Korea-21 Project at Chonnam National
University, Korea.}, Jeong-Sik Kim and Jaedong Choi \thanks{%
Second and third authors would like to acknowledge financial support in part
from Korea Science and Engineering Foundation Grant (R01-2001-00003). }}
\date{}
\maketitle

\begin{quote}
{\bf Abstract.} We obtain certain inequalities involving several intrinsic
invariants namely scalar curvature, Ricci curvature and $k$-Ricci curvature,
and main extrinsic invariant namely squared mean curvature for submanifolds
in a locally conformal almost cosymplectic manifold with pointwise constant $%
\varphi $-sectional curvature. Applying these inequalities we obtain several
inequalities for slant, invariant, anti-invariant and {\em CR}-submanifolds.
The equality cases are also discussed. \medskip


\end{quote}

\section{Introduction}

``To establish simple relationship between the main intrinsic invariants and
the main extrinsic invariants of a submanifold'' is one of the most
fundamental problems in submanifold theory as recalled by B.-Y. Chen (\cite
{chen99}). Scalar curvature and Ricci curvature are among the main intrinsic
invariants, while the squared mean curvature is the main extrinsic invariant
of a submanifold. For more details we refer to \cite{chen00}. \medskip

On the other hand, there is an interesting class of almost contact metric
manifolds which are locally conformal to almost cosymplectic manifolds (\cite
{GY69}). These manifolds are called locally conformal almost cosymplectic
manifolds (\cite{olszak89}). \medskip

In this paper, we continue the study (\cite{TKK-PIAS}) of submanifolds
tangent to the structure vector field $\xi $ in locally conformal almost
cosymplectic manifolds of pointwise constant $\varphi $-sectional curvature
and obtain certain relations of intrinsic invariants, namely scalar, Ricci
and $k$-Ricci curvatures, with the main extrinsic invariant, namely squared
mean curvature for these submanifolds. The paper is organized as follows.
Section~\ref{sect-lcacm} contains a brief introduction to locally conformal
almost cosymplectic manifolds, while in section~\ref{sect-submfd} some
necessary details about different kind of submanifolds are presented. In
section~\ref{sect-scalar}, we obtain inequalities with left hand side
containing scalar curvature and right hand side containing squared mean
curvature for slant, invariant, anti-invariant and {\em CR}-submanifolds in
locally conformal almost cosymplectic manifolds of pointwise constant $%
\varphi $-sectional curvature. The equality cases hold if and only if these
submanifolds are totally geodesic. In section~\ref{sect-ricci}, we obtain an
inequality involving Ricci curvature and squared mean curvature along with
discussion of equality cases. We then apply these inequalities to find
several inequalities for slant, invariant, anti-invariant and {\em CR}%
-submanifolds. In the last section, we find a relationship between the $k$%
-Ricci curvature and the squared mean curvature for slant, invariant,
anti-invariant and {\em CR}-submanifolds.

\section{Locally conformal almost cosymplectic manifolds\label{sect-lcacm}}

Let $\tilde{M}$ be a $\left( 2m+1\right) $-dimensional almost contact
manifold (\cite{blair02}) endowed with an almost contact structure $(\varphi
,\xi ,\eta )$ consisting of a $\left( 1,1\right) $ tensor field $\varphi $,
a vector field $\xi $, and a $1$-form $\eta $ satisfying $\varphi
^{2}=-I+\eta \otimes \xi $ and (one of) $\eta (\xi )=1$, $\varphi \xi =0$, $%
\eta \circ \varphi =0$. The almost contact structure induces a natural
almost complex structure $J$ on the product manifold $\tilde{M}\times {\Bbb R%
}$ defined by $J(X,\lambda d/dt)=(\varphi X-\lambda \xi ,\eta (X)d/dt)$,
where $X$ is tangent to $\tilde{M}$, $t$ the coordinate of ${\Bbb R}$ and $%
\lambda $ a smooth function on $\tilde{M}\times {\Bbb R}$. The almost
contact structure is said to be {\em normal} (\cite{SH61}) if the almost
complex structure $J$ is integrable. Let $\left\langle \,,\right\rangle $ be
a compatible Riemannian metric with $(\varphi ,\xi ,\eta )$, i.e., $%
\left\langle X,Y\right\rangle $ $=$ $\left\langle \varphi X,\varphi
Y\right\rangle $ $+$ $\eta (X)\eta (Y)$ or equivalently, $\Phi (X,Y)\equiv
\left\langle X,\varphi Y\right\rangle =-\left\langle \varphi
X,Y\right\rangle $ along with $\left\langle X,\xi \right\rangle =\eta (X)$
for all $X,Y\in T\tilde{M}$. Then, $(\varphi ,\xi ,\eta ,\left\langle
\,,\right\rangle )$ is an almost contact metric structure on $\tilde{M}$,
and $\tilde{M}$ is an almost contact metric manifold. \medskip

If the fundamental $2$-form $\Phi $ and the $1$-form $\eta $ are closed,
then $\tilde{M}$ is said to be {\em almost cosymplectic manifold }(\cite
{GY69}). A normal almost cosymplectic manifold is {\em cosymplectic} (\cite
{blair02}). An almost contact metric structure is cosymplectic if and only
if $\tilde{\nabla}\varphi =0$, where $\tilde{\nabla}$ is the Levi-Civita
connection of the Riemannian metric $\left\langle ,\right\rangle $. An
example of a manifold which has an almost cosymplectic structure which is
not cosymplectic, can be found in \cite{olszak81}. A conformal change of an
almost contact metric structure is defined by $\varphi ^{*}=\varphi $, $\xi
^{*}=e^{-\rho }\xi $, $\eta ^{*}=e^{\rho }\eta $, $\left\langle
,\right\rangle ^{*}=e^{2\rho }\left\langle ,\right\rangle $, where $\rho $
is a differentiable function. $\tilde{M}$ is said to be a {\em locally
conformal almost cosymplectic manifold} if every point of $\tilde{M}$ has a
neighborhood ${\cal U}$ such that $\left( {\cal U},\varphi ^{*},\xi
^{*},\eta ^{*},\left\langle ,\right\rangle ^{*}\right) $ is almost
cosymplectic for some function $\rho $ on ${\cal U}$. Equivalently, $\tilde{M%
}$ is locally conformal almost cosymplectic manifold if there exists a $1$%
-form $\omega $ such that $d\Phi =2\omega \wedge \Phi $, $d\eta =\omega
\wedge \eta $ and $d\omega =0$ (\cite{olszak89}). \medskip

A plane section $\varrho $ in $T_{p}\tilde{M}$ of an almost contact metric
manifold $\tilde{M}$ is called a $\varphi $-{\em section} if $\varrho \perp
\xi $ and $\varphi \left( \varrho \right) =\varrho $. $\tilde{M}$ is of {\em %
pointwise constant} $\varphi $-{\em sectional curvature} if at each point $%
p\in $ $\tilde{M}$, the sectional curvature $\tilde{K}(\varrho )$ does not
depend on the choice of the $\varphi $-section $\varrho $ of $T_{p}\tilde{M}$%
, and in this case for $p\in \tilde{M}$ and for any $\varphi $-section $%
\varrho $ of $T_{p}\tilde{M}$, the function $c$ defined by $c\left( p\right)
=\tilde{K}(\varrho )$ is called the $\varphi ${\em -sectional curvature} of $%
\tilde{M}$. A locally conformal almost cosymplectic manifold $\tilde{M}$ of
dimension $\geq 5$ is of pointwise constant $\varphi $-sectional curvature
if and only if its curvature tensor $\tilde{R}$ is of the form (\cite
{olszak89})
\begin{eqnarray}
\tilde{R}\left( X,Y\right) Z &=&\frac{c-3f^{2}}{4}\left\{ \left\langle
Y,Z\right\rangle X-\left\langle X,Z\right\rangle Y\right\}  \nonumber \\
&&+\ \frac{c+f^{2}}{4}\left\{ 2\left\langle X,\varphi Y\right\rangle \varphi
Z+\left\langle X,\varphi Z\right\rangle \varphi Y-\left\langle Y,\varphi
Z\right\rangle \varphi X\right\}  \nonumber \\
&&+\ \left( \!\frac{c+f^{2}}{4}\!+\!f^{\prime }\!\right) \!\{\eta (X)\eta
(Z)Y\!-\!\eta (Y)\eta (Z)X\!  \nonumber \\
&&+\ \left\langle X,Z\right\rangle \eta (Y)\xi \!-\!\left\langle
Y,Z\right\rangle \eta (X)\xi \}  \label{curv-tens}
\end{eqnarray}
for all $X,Y,Z\in T\tilde{M}$, where $f$ is the function such that $\omega
=f\eta $, $f^{\prime }=\xi f$; and $c$ is the pointwise $\varphi $-sectional
curvature of $\tilde{M}$.

\section{Submanifolds\label{sect-submfd}}

Let $M$ be an $n$-dimensional Riemannian manifold. The scalar curvature $%
\tau $ at $p$ is given by $\tau =\sum_{i<j}K_{ij}$, where $K_{ij}$ is the
sectional curvature of the plane section spanned by $e_{i}$ and $e_{j}$ at $%
p\in M$ for any orthonormal basis $\{e_{1},\ldots ,e_{n}\}$ for $T_{p}M$.
Now, if $M$ is immersed in an $m$-dimensional Riemannian manifold $\left(
\tilde{M},\left\langle \,,\right\rangle \right) $, then Gauss and Weingarten
formulas are given respectively by $\tilde{\nabla}_{X}Y=\nabla _{X}Y+\sigma
\left( X,Y\right) $ and $\tilde{\nabla}_{X}N=-A_{N}X+\nabla _{X}^{\perp }N$
for all $X,Y\in TM$ and $N\in T^{\perp }M$, where $\tilde{\nabla}$, $\nabla $
and $\nabla ^{\perp }$ are Riemannian, induced Riemannian and induced normal
connections in $\tilde{M}$, $M$ and the normal bundle $T^{\perp }M$ of $M$
respectively, and $\sigma $ is the second fundamental form related to the
shape operator $A_{N}$ in the direction of $N$ by $\left\langle \sigma
\left( X,Y\right) ,N\right\rangle =\left\langle A_{N}X,Y\right\rangle $.
Then, the Gauss equation is
\begin{equation}
\tilde{R}(X,Y,Z,W)=R(X,Y,Z,W)-\left\langle \sigma \left( X,W\right) ,\sigma
\left( Y,Z\right) \right\rangle +\left\langle \sigma \left( X,Z\right)
,\sigma \left( Y,W\right) \right\rangle  \label{gauss-eqn}
\end{equation}
for all $X,Y,Z,W\in TM$, where $\tilde{R}$ and $R$ are the curvature tensors
of $\tilde{M}$ and $M$ respectively. Let $\left\{ e_{1},...,e_{n}\right\} $
be an orthonormal basis of the tangent space $T_{p}M$. The mean curvature
vector $H$ at $p\in M$ is expressed by
\begin{equation}
nH=\hbox{trace}\left( \sigma \right) =\sum_{i=1}^{n}\sigma \left(
e_{i},e_{i}\right) .  \label{mean-curv}
\end{equation}
The submanifold $M$ is {\em totally geodesic} in $\tilde{M}$ if $\sigma =0$,
and {\em minimal} if $H=0$. If $\sigma \left( X,Y\right) $ $=$ $\left\langle
X,Y\right\rangle H$ for all $X,Y\in TM$, then $M$ is {\em totally umbilical}%
. We put
\[
\sigma _{ij}^{r}=g\left( \sigma \left( e_{i},e_{j}\right)
,e_{r}\right) \quad \hbox{and}\quad \left\| \sigma \right\|
^{2}=\sum_{i,j=1}^{n}g\left( \sigma \left( e_{i},e_{j}\right)
,\sigma \left( e_{i},e_{j}\right) \right) ,
\]
where $e_{r}$ belongs to an orthonormal basis $\left\{
e_{n+1},...,e_{m}\right\} $ of the normal space $T_{p}^{\perp }M$.

Let ${\cal L}$ be a $k$\/-plane section of $T_{p}M$ and $X$ a unit vector in
${\cal L}$. We choose an orthonormal basis $\{e_{1},...,e_{k}\}$ of ${\cal L}
$ such that $e_{1}=X.$ The Ricci curvature $\hbox{Ric}_{{\cal L}}$ of ${\cal %
L}$ at $X$ is given by
\begin{equation}
\hbox{Ric}_{{\cal L}}(X)=K_{12}+K_{13}+\cdots +K_{1k},
\label{ricci-L-1}
\end{equation}
where $K_{ij}$ denotes the sectional curvature of the $2$-plane section
spanned by $e_{i},e_{j}$. $\hbox{Ric}_{{\cal L}}(X)$ is called a $k$\/-{\em %
Ricci curvature}. The scalar curvature $\tau $ of the $k$\/-plane section $%
{\cal L}$ is given by
\begin{equation}
\tau \left( {\cal L}\right) =\sum_{1\leq i<j\leq k}K_{ij}.  \label{ricci-L-2}
\end{equation}
For each integer $k,\,2\leq k\leq n$, the Riemannian invariant $\theta _{k}$
on an $n$-dimensional Riemannian manifold $M$ is defined by
\begin{equation}
\theta _{k}\left( p\right) =\left( \frac{1}{k-1}\right) \inf_{{\cal L},X}%
\hbox{Ric}_{{\cal L}}\left( X\right) ,\quad p\in M,
\label{ricci-L-3}
\end{equation}
where ${\cal L}$ runs over all $k$-plane sections in $T_{p}M$ and $X$ runs
over all unit vectors in ${\cal L}$. \medskip

Now, let $M$ be an $n$-dimensional submanifold in an almost contact metric
manifold. For a vector field $X$ in $M$, we put
\[
\varphi X=PX+FX,\qquad PX\in TM,\ FX\in T^{\perp }M\hbox{.}
\]
Thus, $P$ is an endomorphism of the tangent bundle of $M$ and satisfies $%
\left\langle X,PY\right\rangle =-\left\langle PX,Y\right\rangle $ for all $%
X,Y\in TM$. We can define the squared norm of $P$ by
\[
\left\| P\right\| ^{2}\,=\sum_{i,j=1}^{n}\left\langle
e_{i},Pe_{j}\right\rangle ^{2}
\]
for any local orthonormal basis $\{e_{1},e_{2},\ldots ,e_{n}\}$ for $T_{p}M$%
. If the structure vector field $\xi $ is tangential to $M$, then we write
the orthogonal direct decomposition $TM={\cal E}\oplus {\cal E}^{\perp }$,
where ${\cal E}$ is the distribution spanned by $\xi $.

A submanifold $M$ of an almost contact metric manifold with $\xi \in TM$ is
called a {\em semi-invariant submanifold} (\cite{bejancu86}) or a {\em %
contact CR submanifold} (\cite{YK84}) if there exists two differentiable
distributions ${\cal D}$ and ${\cal D}^{\perp }$ on $M$ such that {\bf (i) }$%
TM={\cal D}\oplus {\cal D}^{\perp }\oplus {\cal E}$, {\bf (ii) }the
distribution ${\cal D}$ is invariant by $\varphi $, i.e., $\varphi ({\cal D}%
)={\cal D}$, and {\bf (iii) }the distribution ${\cal D}^{\perp }$ is
anti-invariant by $\varphi $, i.e., $\varphi ({\cal D}^{\perp })\subseteq
T^{\perp }M$.

The submanifold $M$ tangent to $\xi $ is said to be {\em invariant} or {\em %
anti-invariant} (\cite{YK84}) according as $F=0$ or $P=0$. Thus, a {\em CR}%
-submanifold is invariant or anti-invariant according as ${\cal D}^{\perp
}=\left\{ 0\right\} $ or ${\cal D}=\left\{ 0\right\} $. A proper {\em CR}%
-submanifold is neither invariant nor anti-invariant.

For each non zero vector $X\in T_{p}M$, such that $X$ is not proportional to
$\xi _{p}$, we denote the angle between $\varphi X$ and $T_{p}M$ by $\theta
\left( X\right) $. Then $M$ is said to be {\em slant} (\cite{CCFF00},\cite
{lotta96}) if the angle $\theta \left( X\right) $ is constant, i.e., it is
independent of the choice of $p\in M$ and $X\in T_{p}M-\left\{ \xi \right\} $%
. The angle $\theta $ of a slant immersion is called the {\em slant angle}
of the immersion. Invariant and anti-invariant immersions are slant
immersions with slant angle $\theta =0$ and $\theta =\pi /2$ respectively. A
{\em proper} slant immersion is neither invariant nor anti-invariant.

\section{Scalar curvature\label{sect-scalar}}

Let $M$ be an $n$-dimensional submanifold in a locally conformal almost
cosymplectic manifold $\tilde{M}\left( c\right) $ of pointwise constant $%
\varphi $-sectional curvature $c$ such that the structure vector field $\xi $
is tangential to $M$. In view of (\ref{curv-tens}) and (\ref{gauss-eqn}), it
follows that
\begin{eqnarray}
R\left( X,Y,Z,W\right) &=&\frac{c-3f^{2}}{4}\left\{ \left\langle
X,W\right\rangle \left\langle Y,Z\right\rangle -\left\langle
X,Z\right\rangle \left\langle Y,W\right\rangle \right\}  \nonumber \\
&&+\frac{c+f^{2}}{4}\{\left\langle X,PW\right\rangle \left\langle
Y,PZ\right\rangle -\left\langle X,PZ\right\rangle \left\langle
Y,PW\right\rangle  \nonumber \\
&&\qquad -2\left\langle X,PY\right\rangle \left\langle Z,PW\right\rangle \}
\nonumber \\
&&+\left( \frac{c+f^{2}}{4}+f^{\prime }\right) \left\{ \eta (X)\eta
(Z)\left\langle Y,W\right\rangle -\eta (Y)\eta (Z)\left\langle
X,W\right\rangle \right.  \nonumber \\
&&+\left. \left\langle X,Z\right\rangle \eta (Y)\eta \left( W\right)
-\left\langle Y,Z\right\rangle \eta (X)\eta \left( W\right) \right\}
\nonumber \\
&&+\left\langle \sigma (X,W),\sigma (Y,Z)\right\rangle -\left\langle \sigma
(X,Z),\sigma (Y,W)\right\rangle  \label{gauss-eq}
\end{eqnarray}
for all $X,Y,Z,W\in TM$. Thus, the scalar curvature and the mean curvature
of $M$ at $p$ satisfy (\cite{TKK-PIAS})
\begin{eqnarray}
n^{2}\left\| H\right\| ^{2} &=&2\tau +\left\| \sigma \right\| ^{2}-\frac{1}{4%
}n(n-1)\left( c-3f^{2}\right)  \nonumber \\
&&-\frac{3}{4}\left\| P\right\| ^{2}\left( c+f^{2}\right) +2\left(
n-1\right) \left( \frac{c+f^{2}}{4}{+}f^{^{\prime }}\right) .
\label{tau-H-sigma}
\end{eqnarray}
Thus, we are able to state the following

\begin{th}
\label{th-scalar}For an $n$-dimensional submanifold $M$ in a locally
conformal almost cosymplectic manifold $\tilde{M}\left( c\right) $ of
pointwise constant $\varphi $-sectional curvature $c$ such that $\xi \in TM$%
, the following statements are true. \newline
{\bf 1.} We have
\begin{eqnarray}
\tau &\leq &\frac{1}{2}n^{2}\left\| H\right\| ^{2}+\frac{1}{8}n(n-1)\left(
c-3f^{2}\right)  \nonumber \\
&&+\frac{1}{8}\{3\left\| P\right\| ^{2}-2\left( n-1\right) \}\left(
c+f^{2}\right) -\left( n-1\right) f^{\prime }.  \label{scalar-lc}
\end{eqnarray}
{\bf 2.} If $M$ is a $\theta $-slant submanifold, then
\begin{equation}
\tau \leq \frac{1}{2}n^{2}\left\| H\right\| ^{2}+\frac{n-1}{8}%
\{n(c-3f^{2})+(3\cos ^{2}\theta -2)(c+f^{2})-8f^{\prime }\}.
\label{scalar-lc-slant}
\end{equation}
{\bf 3. }If $M$ is an invariant submanifold, then
\begin{equation}
\tau \leq \frac{1}{2}n^{2}\left\| H\right\| ^{2}+\frac{n-1}{8}%
\{(n+1)c-(3n-1)f^{2}-8f^{\prime }\}.  \label{scalar-lc-inv}
\end{equation}
{\bf 4. }If $M$ is an anti-invariant submanifold, then
\begin{equation}
\tau \leq \frac{1}{2}n^{2}\left\| H\right\| ^{2}+\frac{n-1}{8}%
\{(n-2)c-(3n+2)f^{2}-8f^{\prime }\}.  \label{scalar-lc-anti}
\end{equation}
{\bf 5.} If $M$ is a CR-submanifold, then
\begin{eqnarray}
\tau &\leq &\frac{1}{2}n^{2}\left\| H\right\| ^{2}+\frac{1}{8}n(n-1)\left(
c-3f^{2}\right)  \nonumber \\
&&+\frac{1}{8}\{6h-2\left( n-1\right) \}\left( c+f^{2}\right) -\left(
n-1\right) f^{\prime },  \label{scalar-lc-cr}
\end{eqnarray}
where $2h=\dim \left( {\cal D}\right) $. \newline
{\bf 6.} The equality cases of {\em (\ref{scalar-lc})}, {\em (\ref
{scalar-lc-slant})}, {\em (\ref{scalar-lc-inv})}, {\em (\ref{scalar-lc-anti})%
} and {\em (\ref{scalar-lc-cr}) }hold if and only if $M$ is totally geodesic.
\end{th}

\noindent {\bf Proof.} It is easy to verify that an $n$-dimensioanl $\theta $%
-slant submanifold $M$ of an almost contact metric manifold satisfies
\begin{equation}
\left\langle PX,PY\right\rangle =\cos ^{2}\theta \left\langle
\varphi X,\varphi Y\right\rangle \hbox{,\quad }\left\langle
FX,FY\right\rangle =\sin ^{2}\theta \left\langle \varphi
X,\varphi Y\right\rangle  \label{slant-1}
\end{equation}
for all $X,Y\in TM$. Consequently,
\begin{equation}
\left\| P\right\| ^{2}\,=\left( n-1\right) \cos ^{2}\theta
\hbox{.} \label{slant-2}
\end{equation}
If $M$ is a {\em CR}-submanifold, then
\begin{equation}
\left\| P\right\| ^{2}\,=2h=\dim \left( {\cal D}\right) \hbox{.}
\label{cr-1}
\end{equation}
Inequality (\ref{scalar-lc}) follows from (\ref{tau-H-sigma}). Using (\ref
{slant-2}) in (\ref{scalar-lc}), we get (\ref{scalar-lc-slant}). In (\ref
{scalar-lc-slant}), putting $\theta =0$ and $\theta =\pi /2$ we get (\ref
{scalar-lc-inv}) and (\ref{scalar-lc-anti}) respectively. Using (\ref{cr-1})
in (\ref{scalar-lc}), we get (\ref{scalar-lc-cr}). The sixth statement is
obvious in view of (\ref{tau-H-sigma}). $\square $

\section{Ricci curvature\label{sect-ricci}}

For an $n$-dimensional submanifold in an $m$-dimensional Riemannian
manifold, let $\left\{ e_{1},...,e_{n}\right\} $ be a local orthonormal
basis for $T_{p}M$ and $\left\{ e_{n+1},...,e_{m}\right\} $ an orthonormal
basis for the normal space $T_{p}^{\perp }M$. Then, it is easy to verify
that at each point $p\in M$ the squared second fundamental form and the
squared mean curvature satisfy
\begin{eqnarray}
\Vert \sigma \Vert ^{2} &=&\frac{1}{2}n^{2}\Vert H\Vert ^{2}+\frac{1}{2}%
\sum_{r=n+1}^{m}(\sigma _{11}^{r}-\sigma _{22}^{r}-\cdots -\sigma
_{nn}^{r})^{2}  \nonumber \\
&&+\ 2\sum_{r=n+1}^{m}\sum_{j=2}^{n}(\sigma
_{1j}^{r})^{2}-2\sum_{r=n+1}^{m}\sum_{2\leq i<j\leq n}\left( \sigma
_{ii}^{r}\sigma _{jj}^{r}-(\sigma _{ij}^{r})^{2}\right) .  \label{H-sigma}
\end{eqnarray}
The {\em relative null space} of $M$ at a point $p\in M$ is defined by (\cite
{chen99})
\[
N_{p}=\left\{ X\in T_{p}M|\,\sigma (X,Y)=0\ \hbox{for all }Y\in
T_{p}M\right\} .
\]

In \cite{chen99}, B.-Y. Chen established a relationship between Ricci
curvature and the squared mean curvature for a submanifold in a real space
form as follows.

\begin{th}
\label{th-byc}Let $M$ be an $n$-dimensional submanifold in a real space form
$R^{m}\left( c\right) $. Then, \newline
{\bf 1.} For each unit vector $X$ $\in T_{p}M$, we have
\begin{equation}
\Vert H\Vert ^{2}\geq \frac{4}{n^{2}}\left\{ \hbox{Ric}\left(
X\right) -\left( n-1\right) c\right\} .  \label{ricci-byc}
\end{equation}
{\bf 2.} If $H(p)=0$, then a unit vector $X\in T_{p}M$ satisfies the
equality case of {\em (\ref{ricci-byc})} if and only if $X$ lies in the
relative null space $N_{p}$ at $p$. \newline
{\bf 3.} The equality case of {\em (\ref{ricci-byc})}\ holds for all unit
vectors $X$ $\in T_{p}M$, if and only if either $p$\ is a totally geodesic
point or $n=2$ and $p$ is a totally umbilical point.
\end{th}

In this section, we find similar results for several kind of submanifolds in
a locally conformal almost cosymplectic manifold.

\begin{th}
\label{th-ric-xi-hor}Let $M$ be an $n$-dimensional submanifold in a $(2m+1)$%
-dimen\-sional locally conformal almost cosymplectic manifold $\tilde{M}%
\left( c\right) $ of pointwise constant $\varphi $-sectional curvature $c$
tangential to the structure vector field $\xi $. Then, the following
statements are true. \newline
{\bf (i)} For each unit vector $X$ $\in T_{p}M$, we have
\begin{eqnarray}
\hbox{Ric}\left( X\right) &\leq &\frac{1}{4}\left\{ n^{2}\Vert
H\Vert
^{2}+\left( n-1\right) \left( c-3f^{2}\right) \right.  \nonumber \\
&&\left. +\left( 3\left\| PX\right\| ^{2}-\left( n-2\right) \eta \left(
X\right) ^{2}-1\right) \left( c+f^{2}\right) \right\}  \nonumber \\
&&-\left( 1\!+\left( n-2\right) \eta \left( X\right) ^{2}\right) f^{^{\prime
}}.  \label{ricci-1}
\end{eqnarray}
{\bf (ii)} For $H(p)=0$, a unit vector $X\in T_{p}M$ satisfies the equality
case of {\em (\ref{ricci-1})} if and only if $X$ belongs to the relative
null space $N_{p}$ at $p$. \newline
{\bf (iii)} The equality case of {\em (\ref{ricci-1})}\ holds identically
for all unit vectors $X$ $\in T_{p}M$, if and only if either $n=2$\ and $p$\
is a totally umbilical point or $p$\ is a totally geodesic point.
\end{th}

\noindent {\bf Proof. }We choose an orthonormal basis $\left\{
e_{1},...,e_{n},e_{n+1},...,e_{2m+1}\right\} $ such that $e_{1},...,e_{n}$
are tangential to $M$ at $p$. From (\ref{gauss-eq}), we get
\begin{eqnarray*}
K_{ij} &=&\sum_{r=n+1}^{2m+1}\left( \sigma _{ii}^{r}\sigma _{jj}^{r}-(\sigma
_{ij}^{r})^{2}\right) +\frac{c-3f^{2}}{4}+\frac{3\left( c+f^{2}\right) }{4}%
\left\langle e_{i},Pe_{j}\right\rangle ^{2} \\
&&-\left( \frac{c+f^{2}}{4}{+}f^{^{\prime }}\right) \left( \eta \left(
e_{i}\right) ^{2}+\eta \left( e_{j}\right) ^{2}\right) ,
\end{eqnarray*}
which gives
\begin{eqnarray}
\sum_{2\leq i<j\leq n}K_{ij} &=&\sum_{r=n+1}^{2m+1}\sum_{2\leq i<j\leq
n}\left( \sigma _{ii}^{r}\sigma _{jj}^{r}-(\sigma _{ij}^{r})^{2}\right) +%
\frac{1}{8}\left( n-1\right) (n-2)\left( c-3f^{2}\right)  \nonumber \\
&&+\frac{1}{8}\left( 3\left\| P\right\| ^{2}-6\left\| Pe_{1}\right\|
^{2}\right) \left( c+f^{2}\right)  \nonumber \\
&&-\left( n-2\right) \left( \frac{c+f^{2}}{4}{+}f^{^{\prime }}\right) \left(
1-\eta \left( e_{1}\right) ^{2}\right) .  \label{ricci-2}
\end{eqnarray}
From (\ref{tau-H-sigma}) and (\ref{H-sigma}), we get
\begin{eqnarray}
\frac{1}{4}n^{2}\Vert H\Vert ^{2} &=&\tau -\frac{1}{8}n(n-1)\left(
c-3f^{2}\right) -\frac{1}{8}\left( 3\left\| P\right\| ^{2}-2\left(
n-1\right) \right) \left( c+f^{2}\right)  \nonumber \\
&&+\ \left( n-1\right) f^{\prime }+\ \frac{1}{4}\sum_{r=n+1}^{2m+1}(\sigma
_{11}^{r}-\sigma _{22}^{r}-\cdots -\sigma _{nn}^{r})^{2}  \nonumber \\
&&+\ \sum_{r=n+1}^{2m+1}\sum_{j=2}^{n}(\sigma
_{1j}^{r})^{2}-\sum_{r=n+1}^{2m+1}\sum_{2\leq i<j\leq n}\left( \sigma
_{ii}^{r}\sigma _{jj}^{r}-(\sigma _{ij}^{r})^{2}\right) ,  \label{ricci-3}
\end{eqnarray}
which in view of (\ref{ricci-2}) provides
\begin{eqnarray}
\hbox{Ric}\left( e_{1}\right) &=&\frac{1}{4}\left\{ n^{2}\Vert
H\Vert
^{2}+\left( n-1\right) \left( c-3f^{2}\right) \right.  \nonumber \\
&&\left. +\left( 3\left\| Pe_{1}\right\| ^{2}-\left( n-2\right) \eta \left(
e_{1}\right) ^{2}-1\right) \left( c+f^{2}\right) \right\}  \nonumber \\
&&-\left( 1+\left( n-2\right) \eta \left( e_{1}\right) ^{2}\right)
f^{^{\prime }}  \nonumber \\
&&-\frac{1}{4}\!\sum_{r=n+1}^{2m+1}\!(\sigma _{11}^{r}-\sigma
_{22}^{r}-\cdots -\sigma
_{nn}^{r})^{2}\!-\!\sum_{r=n+1}^{2m+1}\sum_{j=2}^{n}(\sigma _{1j}^{r})^{2}.
\label{ricci-4}
\end{eqnarray}
Since $e_{1}=X$ can be chosen to be any arbitrary unit vector in $T_{p}M$,
therefore the above equation implies (\ref{ricci-1}).

In view of (\ref{ricci-4}), the equality case of (\ref{ricci-1}) is valid if
and only if
\begin{equation}
\sigma _{12}^{r}=\cdots =\sigma _{1n}^{r}=0\ \hbox{and }\sigma
_{11}^{r}=\sigma _{22}^{r}+\cdots +\sigma _{nn}^{r},\quad r\in
\left\{ n+1,\ldots ,2m+1\right\} .  \label{ricci-5}
\end{equation}
If $H(p)=0$, (\ref{ricci-5}) implies that $e_{1}=X$ belongs to the relative
null space $N_{p}$ at $p$. Conversely, if $e_{1}=X$ lies in the relative
null space, then (\ref{ricci-5}) holds because $H(p)=0$ is assumed. This
proves statement {\bf (ii)}.

Now, we prove {\bf (iii)}. Assume that the equality case of (\ref{ricci-1})
for all unit tangent vectors to $M$ at $p\in M$ is true. Then, in view of (%
\ref{ricci-4}), for each $r\in \left\{ n+1,\ldots ,2m+1\right\} $, we have
\begin{equation}
\sigma _{ij}^{r}=0,\quad i\neq j,  \label{ricci-6}
\end{equation}
\begin{equation}
2\sigma _{ii}^{r}=\sigma _{11}^{r}+\cdots +\sigma _{nn}^{r},\quad i\in
\left\{ 1,...,n\right\} .  \label{ricci-7}
\end{equation}
Thus, we have two cases, namely either $n=2$ or $n\neq 2$. In the first case
$p$ is a totally umbilical point, while in the second case $p$ is a totally
geodesic point. The converse is obvious. $\square $\medskip

The above theorem implies the following results for slant, invariant and
anti-invariant submanifolds.

\begin{th}
Let $M$ be an $n$-dimensional submanifold in a locally conformal almost
cosymplectic manifold $\tilde{M}\left( c\right) $ of pointwise constant $%
\varphi $-sectional curvature $c$ such that $\xi \in TM$. Then, the
following statements are true. \newline
{\bf 1.} If $M$ is $\theta $-slant, then for each unit vector $X$ $\in
T_{p}M $, we have
\begin{eqnarray}
\hbox{Ric}\left( X\right) &\leq &\frac{1}{4}\left\{ n^{2}\Vert
H\Vert
^{2}+\left( n-1\right) \left( c-3f^{2}\right) \right.  \nonumber \\
&&\left. +\left( 3\cos ^{2}\theta -\left( n+3\cos ^{2}\theta -2\right) \eta
\left( X\right) ^{2}-1\right) \left( c+f^{2}\right) \right\}  \nonumber \\
&&-\left( 1+\left( n-2\right) \eta \left( X\right) ^{2}\right) f^{^{\prime
}}.  \label{ricci-slant}
\end{eqnarray}
{\bf 2.} If $M$ is invariant, then for each unit vector $X$ $\in T_{p}M$, we
have
\begin{eqnarray}
\hbox{Ric}\left( X\right) &\leq &\frac{1}{4}\{n^{2}\Vert H\Vert
^{2}+(n-1)(c-3f^{2})+(2-(n+1)\eta (X)^{2})(c+f^{2})\}  \nonumber \\
&&-\left( 1+\left( n-2\right) \eta \left( X\right) ^{2}\right) f^{^{\prime
}}.  \label{ricci-inv}
\end{eqnarray}
{\bf 3.} If $M$ is anti-invariant, then for each unit vector $X$ $\in T_{p}M$%
, we have
\begin{equation}
\hbox{Ric}\!\left( X\right) \!\leq \!\frac{1}{4}\{n^{2}\Vert
H\Vert ^{2}\!+\!(n\!-\!1)(c\!-\!3f^{2})\!-\!(1\!+\!(n\!-\!2)\eta
(X)^{2})(c\!+\!f^{2}\!+\!4f^{\prime })\}.  \label{ricci-anti}
\end{equation}
{\bf 4.} If $H(p)=0$, a unit vector $X\in T_{p}M$ satisfies the equality
case of {\em (\ref{ricci-slant})}, {\em (\ref{ricci-inv})} and {\em (\ref
{ricci-anti})} if and only if $X\in N_{p}$. \newline
{\bf 5.} The equality case of {\em (\ref{ricci-slant})}, {\em (\ref
{ricci-inv})} and {\em (\ref{ricci-anti})} holds identically for all unit
vectors $X$ $\in T_{p}M$, if and only if either $n=2$\ and $p$\ is a totally
umbilical point or $p$\ is a totally geodesic point.
\end{th}

\noindent {\bf Proof. }In a $\theta $-slant submanifold, for a unit vector $%
X\in T_{p}M$, in view of (\ref{slant-1}), for a unit vector $X\in T_{p}M$,
we get
\[
\left\| PX\right\| ^{2}=\left\langle PX,PX\right\rangle =\cos ^{2}\theta
\left( 1-\eta \left( X\right) ^{2}\right) .
\]
Using this in (\ref{ricci-1}), we get (\ref{ricci-slant}). Putting $\theta
=0 $ and $\theta =\pi /2$ in (\ref{ricci-slant}), we have (\ref{ricci-inv})
and (\ref{ricci-anti}) respectively. Fourth and fifth statements are
obvious. $\square $ \medskip

We also have the following

\begin{cor}
Let $M$ be an $n$-dimensional CR-submanifold in a locally conformal almost
cosymplectic manifold $\tilde{M}\left( c\right) $ of pointwise constant $%
\varphi $-sectional curvature $c$. Then, the following statements are true.
\newline
{\bf 1.} For each unit vector $X$ $\in {\cal D}$, we have
\begin{equation}
4\hbox{Ric}\left( X\right) \leq n^{2}\Vert H\Vert
^{2}+(n+1)c-\left( 3n-5\right) f^{2}-4f^{^{\prime }}.
\label{ricci-cr-1}
\end{equation}
{\bf 2.} For each unit vector $X$ $\in {\cal D}^{\bot }$, we have
\begin{equation}
4\hbox{Ric}\left( X\right) \leq n^{2}\Vert H\Vert ^{2}+\left(
n-2\right) c-\left( 3n-2\right) f^{2}-4f^{\prime }.
\label{ricci-cr-2}
\end{equation}
{\bf 3.} If $H(p)=0$, a unit vector $X\in {\cal D}$ {\em (}resp. ${\cal D}%
^{\bot }${\em )} satisfies the equality case of {\em (\ref{ricci-cr-1})}
{\em (}resp. {\em (\ref{ricci-cr-2}))} if and only if $X\in N_{p}$.
\end{cor}

\section{$k$-Ricci curvature\label{chen-5}}

In this section, we prove a relationship between the $k$\/-Ricci curvature
and the squared mean curvature for submanifolds in a locally conformal
almost cosymplectic manifold $\tilde{M}\left( c\right) $ of pointwise
constant $\varphi $-sectional curvature $c$ such that $\xi \in TM$.

\begin{th}
\label{th-k-ricci-1}Let $M$ be an $n$-dimensional submanifold in a $\left(
2m+1\right) $-dimen\-sioanl locally conformal almost cosymplectic manifold $%
\tilde{M}\left( c\right) $ of pointwise constant $\varphi $-sectional
curvature $c$ such that $\xi \in TM$. Then we have
\begin{equation}
\left\| H\right\| ^{2}\geq \frac{2\tau }{n(n-1)}-\frac{c-3f^{2}}{4}-\frac{%
3\left\| P\right\| ^{2}\left( c+f^{2}\right) }{4n(n-1)}+\frac{2}{n}\left(
\frac{c+f^{2}}{4}{+}f^{^{\prime }}\right) .  \label{kricci}
\end{equation}
\end{th}

\noindent {\bf Proof. }Choose an orthonormal basis $%
\{e_{1},...,e_{n},e_{n+1},$ $...,e_{2m+1}\}$ at $p$ such that $e_{n+1}$ is
parallel to the mean curvature vector $H(p)$ and $e_{1},...,e_{n}$
diagonalize the shape operator $A_{n+1}$. Then the shape operators take the
forms
\begin{equation}
A_{n+1}=\left(
\begin{array}{cccc}
a_{1} & 0 & \cdots & 0 \\
0 & a_{2} & \cdots & 0 \\
\vdots & \vdots & \ddots & \vdots \\
0 & 0 & \cdots & a_{n}
\end{array}
\right) ,  \label{kricci-1}
\end{equation}
\begin{equation}
A_{r}=\left( h_{ij}^{r}\right) ,\hbox{ }i,j=1,...,n;\hbox{ }%
r=n+2,...,2m+1,\quad
\hbox{trace}\,A_{r}=\sum_{i=1}^{n}h_{ii}^{r}=0. \label{kricci-2}
\end{equation}
From (\ref{tau-H-sigma}), we get
\begin{eqnarray}
n^{2}\left\| H\right\| ^{2} &=&2\tau
+\sum_{i=1}^{n}a_{i}^{2}+\sum_{r=n+2}^{2m}\sum_{i,j=1}^{n}(h_{ij}^{r})^{2}-%
\frac{1}{4}n(n-1)\left( c-3f^{2}\right)  \nonumber \\
&&-\frac{3}{4}\left\| P\right\| ^{2}\left( c+f^{2}\right) +2\left(
n-1\right) \left( \frac{c+f^{2}}{4}{+}f^{^{\prime }}\right) .
\label{kricci-3}
\end{eqnarray}
Since
\[
0\leq
\sum_{i<j}(a_{i}-a_{j})^{2}=(n-1)\sum_{i}a_{i}^{2}-2\sum_{i<j}a_{i}a_{j},
\]
therefore, we get
\[
n^{2}\left\| H\right\| ^{2}=\left( \sum_{i=1}^{n}a_{i}\right)
^{2}=\sum_{i=1}^{n}a_{i}^{2}+2\sum_{i<j}a_{i}a_{j}\leq
n\sum_{i=1}^{n}a_{i}^{2},
\]
which implies
\begin{equation}
\sum_{i=1}^{n}a_{i}^{2}\geq n\left\| H\right\| ^{2}.  \label{kricci-4}
\end{equation}
From (\ref{kricci-3}) and (\ref{kricci-4}), we obtain
\begin{eqnarray}
n^{2}\left\| H\right\| ^{2} &\geq &2\tau +n\left\| H\right\| ^{2}-\frac{1}{4}%
n(n-1)\left( c-3f^{2}\right)  \nonumber \\
&&-\frac{3}{4}\left\| P\right\| ^{2}\left( c+f^{2}\right) +2\left(
n-1\right) \left( \frac{c+f^{2}}{4}{+}f^{^{\prime }}\right) ,
\label{kricci-5}
\end{eqnarray}
which gives (\ref{kricci}). $\square $ \medskip

Next, we prove the following

\begin{th}
\label{th-k-ricci-2}Let $M$ be an $n$-dimensional submanifold in a locally
conformal almost cosymplectic manifold $\tilde{M}\left( c\right) $ of
pointwise constant $\varphi $-sectional curvature $c$ such that $\xi \in TM$%
. Then, for each integer $k,$ $2\leq k\leq n$, and every point $p\in M$, we
have
\begin{equation}
\left\| H\right\| ^{2}\geq \theta _{k}(p)-\frac{c-3f^{2}}{4}-\frac{3\left\|
P\right\| ^{2}\left( c+f^{2}\right) }{4n(n-1)}+\frac{2}{n}\left( \frac{%
c+f^{2}}{4}{+}f^{^{\prime }}\right) .  \label{kricci'}
\end{equation}
\end{th}

\noindent {\bf Proof. }Let $\{e_{1},...,e_{n}\}$ be an orthonormal basis of $%
T_{p}M$. We denote by ${\cal L}_{i_{1}...i_{k}}$ the $k$-plane section
spanned by $e_{i_{1}},...,e_{i_{k}}.$ From (\ref{ricci-L-1}) and (\ref
{ricci-L-2}), it follows that
\begin{equation}
\tau ({\cal L}_{i_{1}...i_{k}})=\frac{1}{2}\sum_{i\in \{i_{1},...,i_{k}\}}%
\hbox{Ric}_{{\cal L}_{i_{1}...i_{k}}}(e_{i}),  \label{kricci-6}
\end{equation}
\begin{equation}
\tau (p)=\frac{1}{C_{k-2}^{n-2}}\sum_{1\leq i_{1}<\cdots <i_{k}\leq n}\tau (%
{\cal L}_{i_{1}...i_{k}}).  \label{kricci-7}
\end{equation}
Combining (\ref{ricci-L-3}), (\ref{kricci-6}) and (\ref{kricci-7}), we
obtain
\begin{equation}
\tau (p)\geq \frac{n(n-1)}{2}\theta _{k}(p).  \label{kricci-8}
\end{equation}
From (\ref{kricci}) and (\ref{kricci-8}) we get (\ref{kricci'}). $\square $
\medskip

As an application, we have the following Corollary. \medskip

\begin{cor}
\label{cor-kricci}Let $M$ be an $n$-dimensional submanifold in a locally
conformal almost cosymplectic manifold $\tilde{M}\left( c\right) $ of
pointwise constant $\varphi $-sectional curvature $c$ such that $\xi \in TM$%
, and let $k$ be any integer such that $2\leq k\leq n$. Then at each point $%
p\in M$, we have the following statements. \newline
{\bf 1.} If $M$ is $\theta $-slant, then
\begin{equation}
\left\| H\right\| ^{2}\geq \theta _{k}-\frac{c-3f^{2}}{4}-\frac{3\left(
c+f^{2}\right) \cos ^{2}\theta }{4n}+\frac{2}{n}\left( \frac{c+f^{2}}{4}{+}%
f^{^{\prime }}\right) .
\end{equation}
\newline
{\bf 2.} If $M$ is invariant, then
\begin{equation}
\left\| H\right\| ^{2}\geq \theta _{k}-\frac{c-3f^{2}}{4}-\frac{3\left(
c+f^{2}\right) }{4n}+\frac{2}{n}\left( \frac{c+f^{2}}{4}{+}f^{^{\prime
}}\right) .
\end{equation}
\newline
{\bf 3.} If $M$ is anti-invariant, then
\begin{equation}
\left\| H\right\| ^{2}\geq \theta _{k}-\frac{c-3f^{2}}{4}+\frac{2}{n}\left(
\frac{c+f^{2}}{4}{+}f^{^{\prime }}\right) .
\end{equation}
\newline
{\bf 4.} If $M$ is a $CR$\/-submanifold, then
\begin{equation}
\left\| H\right\| ^{2}\geq \theta _{k}-\frac{c-3f^{2}}{4}-\frac{6h\left(
c+f^{2}\right) }{4n\left( n-1\right) }+\frac{2}{n}\left( \frac{c+f^{2}}{4}{+}%
f^{^{\prime }}\right) ,
\end{equation}
where $2h=\dim \left( {\cal D}\right) $.
\end{cor}

\noindent Department of Mathematics and Astronomy, \newline
Lucknow University, \newline
Lucknow 226 007, India\newline
email:\quad {\tt mm\_tripathi@hotmail.com}\bigskip \newline
Department of Mathematics Education, \newline
Sunchon National University, \newline
Chonnam 540-742, Korea\newline
email:\quad {\tt jskim01@hanmir.com}\bigskip \newline
Department of Mathematics,\newline
Airforce Academy, \newline
P. O. Box 335-2 Ssangsu, Namil,\newline
Chungwon, Chungbuk, 363-849, Korea\newline
e-mail : {\tt jdong@afa.ac.kr}\newpage

\thispagestyle{empty}$\displaystyle$

\end{document}